\documentclass{article}

\usepackage{graphicx}

\def\abstract#1{\vskip 7mm 
        \begin{center}{\large Abstract}\par \smallskip
                \begin{minipage}[c]{12cm}
                        \small #1
                \end{minipage}
        \end{center}
}
\def\title#1{\begin{center}{\Large\bf #1}\end{center}}
\def\author#1{\vskip 5mm \begin{center}{#1}\end{center}}
\def\address#1{\begin{center}{\it #1}\end{center}}
\makeatletter
\@ifundefined{lesssim}{}{}
\@ifundefined{gtrsim}{}{}
\def\vereq#1#2{\lower3pt\vbox{\baselineskip1.5pt \lineskip1.5pt
\ialign{$\m@th#1\hfill##\hfil$\crcr#2\crcr\sim\crcr}}}
\makeatother

\begin{document}

\title{%
  Cosmology of multigravity
}
\author{%
  Teruki Hanada\footnote{E-mail:k004wa@yamaguchi-u.ac.jp},
  Kazuhiko Shinoda\footnote{E-mail:k009vc@yamaguchi-u.ac.jp}
  and
  Kiyoshi Shiraishi\footnote{E-mail:shiraish@yamaguchi-u.ac.jp}
}
\address{%
  Graduate School of Science and Engineering, Yamaguchi University,
   Yoshida, Yamaguchi-shi, Yamaguchi 753--8512, Japan
}

\abstract{
We have constructed a nonlinear multi-graviton theory. An application of
this theory to cosmology is discussed. We found that scale
factors in a solution for this theory repeat acceleration and 
deceleration. }

\section{Introduction}
We have constructed a nonlinear multi-graviton theory \cite{ref1} so
far.\footnote{For related work, see~\cite{ref3.6,MG}.} The features of our
model are following: (i) Gravitons as the fluctuation from Minkowski
spacetime have a Fierz-Pauli (FP) type mass
\cite{ref1.5}. (ii) This model based on dimensional deconstruction
method. So, we can tune the mass spectrum more easily than Kaluza-Klein
theory. (iii) The mass term has a reflection symmetry at each vertex and
a exchange symmetry at each edge of a graph.

In this article, beginning with graph theory, dimensional
deconstruction
\cite{ref2,ref3} and our model are reviewed (see also
\cite{ref3.6,ref3.5}).  Next, we consider the vacuum cosmological
solutions of the case with the four-site star graph and the four-site path
graph. Finally, we summarize our work and remark about the outlook.

\section{A review of graph theory and dimensional deconstruction}
\subsection{Graph theory}
We consider the matrix representation of 
graph theory.\footnote{Please see \cite{ref4} for a review of application of graph theory to field theory.}
 A graph $G$ is a pair of $V$ and $E$, where $V$
is a set of vertices while $E$ is a set of edges. An edge connects two vertices; two vertices located at the
ends of an edge $e$ are denoted as $o(e)$ and $t(e)$. Then, we introduce two matrices, an incidence matrix
and a graph Laplacian, associated with a specific graph. The incidence matrix represents the condition
of connection or structure of a graph, and the graph Laplacian $\Delta$
can be obtained by $EE^T$ . By use of these matrices, a quadratic form of
vectors $a^T\Delta a(= a^T E E^T a)$ can be written as a sum of
$(a_i\Delta_{ij} a_j)$. If all $a_i (i = 1, 2, . . . ,\sharp V )$,
components of $a$, take the same value, $E^Ta = 0$ and then
$a = 0$.

\subsection{Dimensional deconstruction}
It is assumed that we put fields on vertices or edges. An idea that there are four dimensional fields
on the sites (vertices) and links (edges), dubbed as dimensional
deconstruction, is introduced by Arkani-Hamed {\it et al.}. In this
scheme, the square of mass matrix is proportional to the Laplacian of the
associated graph. In the case of a cycle graph (a `closed circuit') with
$N$ sites $(C_N)$, when $N$ becomes large, the model on the graph
corresponds to the five-dimensional theory with $S^1$ compactification.
In other words, the mass scale of the model $f$ over $N$ corresponds to
the inverse of the compactification radius $L/(2\pi)$:
\[
M_\ell^2=4f^2(\sin\pi\ell/N)^2\quad \rightarrow \quad M_\ell^2=(2\pi \ell/L)^2\qquad (f/N\to 1/L).
\]
For a cycle graph, the linear graviton model presented in the previous
work~\cite{ref1} coincides with the FP model proposed in~\cite{ref1.5}.
The model is a most general linear graviton theory on a generic graph.

\section{Nonlinear multi-graviton theory on a graph}
We consider a nonlinear multigravity on a graph. Following Nibbelink
{\it et al.} \cite{ref5,ref6}, we introduce the  important `tool':
\[\langle ABCD \rangle\equiv -\varepsilon_{abcd}\varepsilon^{\mu\nu\rho\sigma}A^a_\mu B^b_\nu C^c_\rho D^d_\sigma,\]
where $\varepsilon$ is the totally antisymmetric tensor. Using this tool, we have the Einstein-Hilbert term replacing 
$A$ and $B$ by vierbeins and $C$ and $D$ by a curvature 2-form. 
In addition, because fourth power of vierbein in the angle 
brackets is equal to the determinant of vierbeins,
the use of this tool illuminates that the Einstein-Hilbert term and the
cosmological term  have the same structure.

We assume the following term for each edge of a graph form;
\[\langle (e_1e_1 -e_2e_2)^2 \rangle,\]
where $e_1$ and $e_2$ are vierbeins at two ends of one edge. This term has a reflection symmetry $e\leftrightarrow -e$ 
at each vertex and an exchange symmetry $e_1 \leftrightarrow e_2$ at each
edge.

In the weak-field limit, $e_1=\eta +f_1, e_2=\eta+ f_2$,
\[\langle(e_1e_1-e_2e_2)^2\rangle=
8\left(\left(\left[f_1\right]-\left[f_2\right]\right)^2
-\left[\left(f_1-f_2\right)^2\right]\right)+O(f^3),\]
where $\eta$ is Minkowski metric, and $[f]={\rm tr}f$ for notational simplicity. This quadratic term corresponds to 
the FP mass term
\footnote{It is known that the asymmetric part of $f$ can be omitted. \cite{ref7}}
. On the other hand, $\frac{1}{2}|e|R$ contains the kinetic terms of a graviton in the leading order.

Therefore, in the case of the tree graph, we have the nonlinear Lagrangian of multi-graviton theory 
without higher derivertive and non-local terms,
\[L_0=\frac{1}{2}\exp\Phi\sum_{v\in
V}\left|e^v\right|R^v+\frac{M^2}{24}\sum_{e\in E}
\left\langle\left(e_{o(e)}e_{o(e)}-e_{t(e)}e_{t(e)}\right)^2\right\rangle,\]
where $M^2\equiv 3 m^2/2$ and we assume $\phi_1=\phi_2=\cdots
=\phi_N=\Phi$.

\section{Vacuum cosmological solution}
Now we consider two vacuum cosmological models, based on the four-site star graph and the four-site path graph.
We assume the following metric;
\[g_{\mu\nu}dx^\mu dx^\nu=-e^{\Phi(t)}dt^2+e^{\Phi(t)+2a_i(t)}(dr^2+r^2d\Omega^2)\]
where $\Phi(t)$ is a scalar field and $a_i(i=1,\cdots, 4)$ are scale
factors. Then, 
\[\left\langle (e_ie_i-e_je_j)^2\right\rangle=e^{-2\Phi(t)}(e^{2a_i(t)}-e^{2a_j(t)})^2.\]

In the case of the star graph, the Lagrangian is following;
\[L=\frac{1}{2}\exp\Phi(t)\sum_{i=1}^4|e^i|R^i+\frac{M^2}{24}\sum_{i=2}^4\left\langle
(e_1e_1-e_ie_i)^2\right\rangle,\] where, $a_1$ is on the center of the
graph. On the other hand, the Lagrangian of the case of the path graph is
\[L=\frac{1}{2}\exp\Phi(t)\sum_{i=1}^4|e^i|R^i+\frac{M^2}{24}\sum_{i<j}\left\langle
(e_ie_i-e_je_j)^2\right\rangle,\] where, $a_1$ and $a_4$ are on each end
of the graph.

We show the results of numerical calculations of the two models
 on the same initial conditions in Figure 1 and Figure 2. 
Both scalar fields behave similarly and each scale factor 
repeats the increase and the decrease. However, the scalar field in the star 
graph case changes slightly slower 
than the other. The oscillation of the scale factors in the path graph
case  include more modes of different frequencies than that of the scale
factors in the star graph case.

The star graph model has more symmetries than the path graph model. 
A lot of modes in the star graph case degenerate. 
In the path graph case, increase of the number of sites gives the more complicated behaviors of the scale factors.
On the other hand, in the star graph case, the symmetries are preserved if the number of sites are inceased. 
Therefore, the behaviors of the scale factors are the same as the four-site model, essentially.
\section{Summary and prospects}
We considered the four-site star and path graph model
and found that vacuum cosmological solutions with the scale factors 
show the repeated accelerating and decelerating expansions.
The differences between these two models were discussed from a viewpoint about symmetries.

As the future works, we should investigate the case that the matter fields exist. We also should investigate 
the models based on arbitrary tree graphs.

\section*{Acknowledgements}
The authors would like to thank N. Kan and K. Kobayashi for useful comments, and also the organizers of JGRG18.

\begin{figure}[htbp]
\begin{minipage}{0.33\hsize}
\includegraphics[width=\hsize,keepaspectratio,clip]{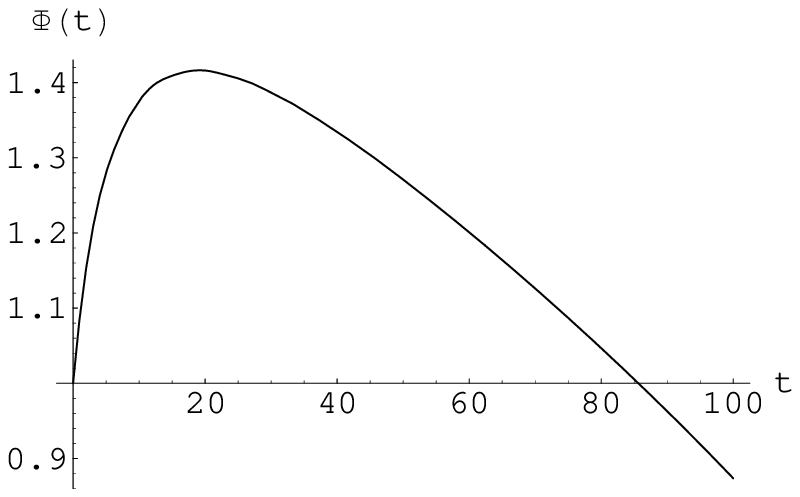}
\end{minipage}
\begin{minipage}{0.33\hsize}
\includegraphics[width=\hsize,keepaspectratio,clip]{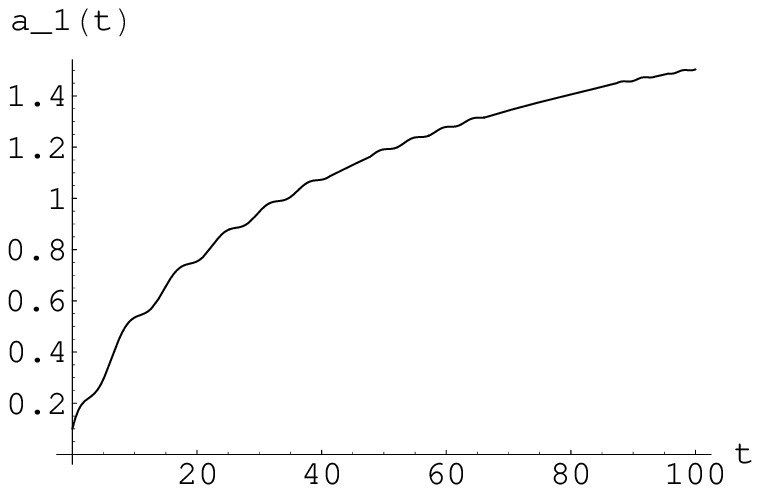}
\end{minipage}
\begin{minipage}{0.33\hsize}
\includegraphics[width=\hsize,keepaspectratio,clip]{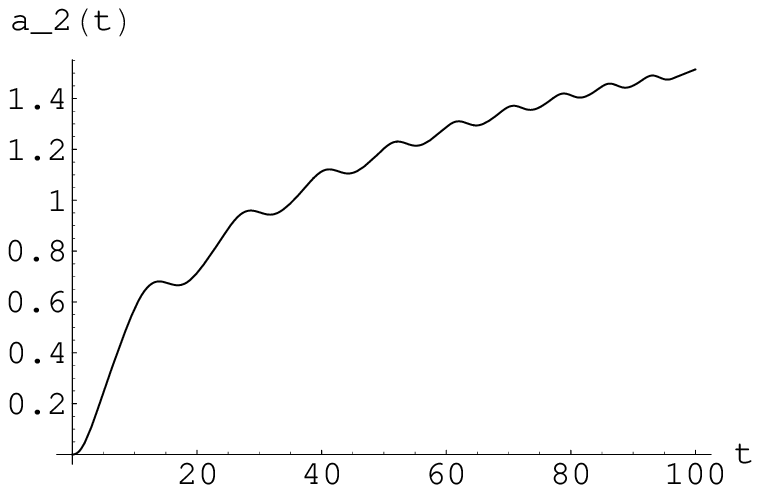}
\end{minipage}\\ \vspace*{.2cm}
\begin{minipage}{0.33\hsize}
\includegraphics[width=\hsize,keepaspectratio,clip]{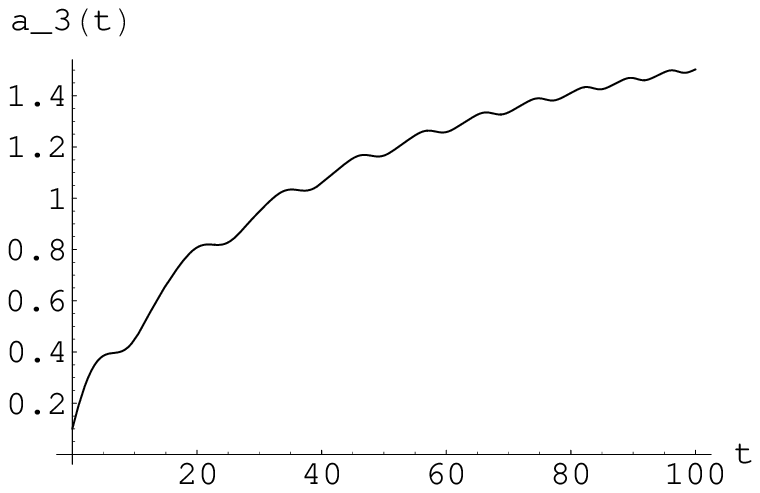}
\end{minipage}
\begin{minipage}{0.33\hsize}
\includegraphics[width=\hsize,keepaspectratio,clip]{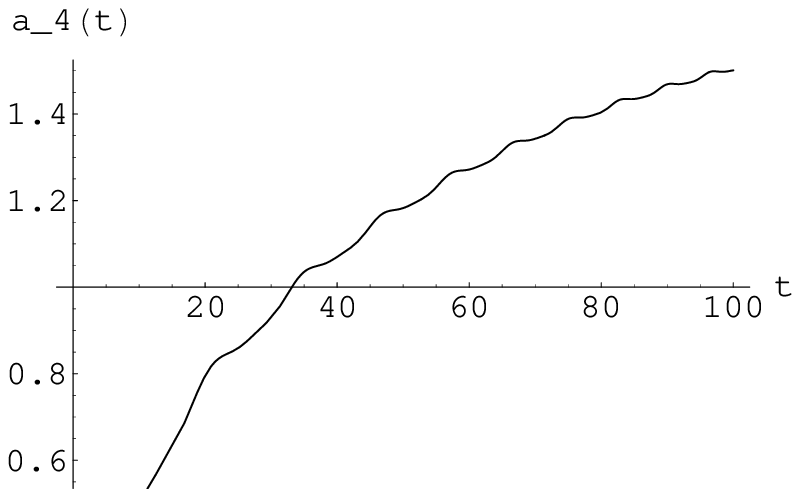}
\end{minipage}
\begin{center}
\caption{A numerical solutions of the scale factors in the case of the four-site star graph. }
\end{center}
\end{figure}

\begin{figure}[htbp]
\begin{minipage}{0.33\hsize}
\includegraphics[width=\hsize,keepaspectratio,clip]{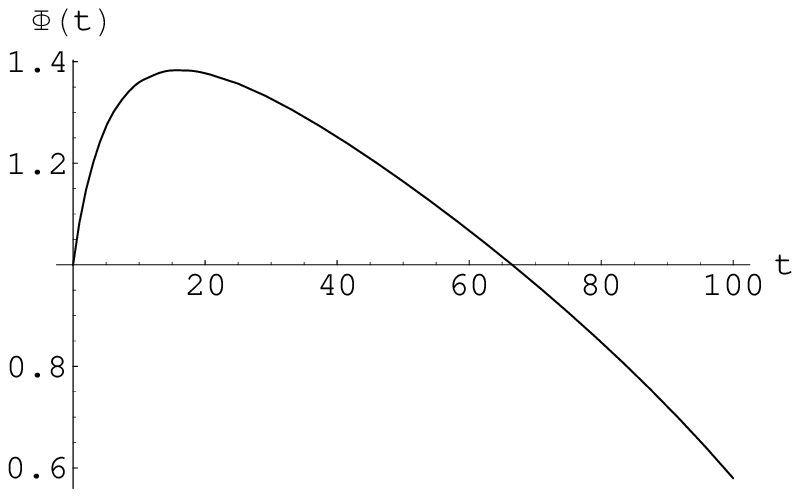}
\end{minipage}
\begin{minipage}{0.33\hsize}
\includegraphics[width=\hsize,keepaspectratio,clip]{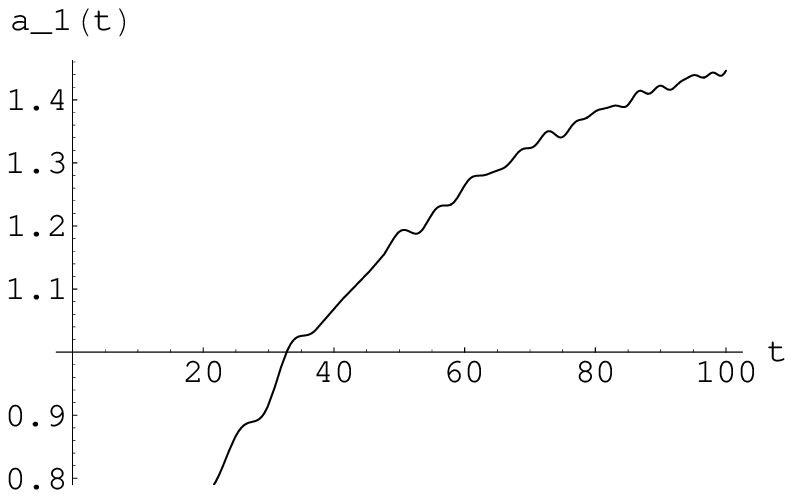}
\end{minipage}
\begin{minipage}{0.33\hsize}
\includegraphics[width=\hsize,keepaspectratio,clip]{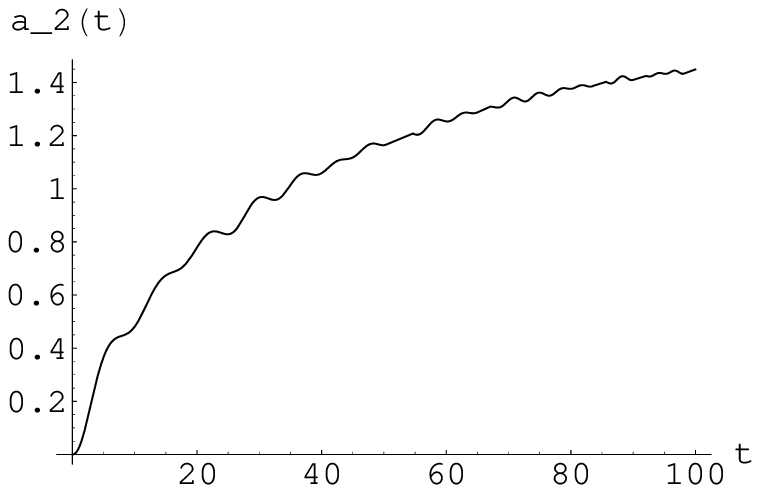}
\end{minipage}
\\ \\
\begin{minipage}{0.33\hsize}
\includegraphics[width=\hsize,keepaspectratio,clip]{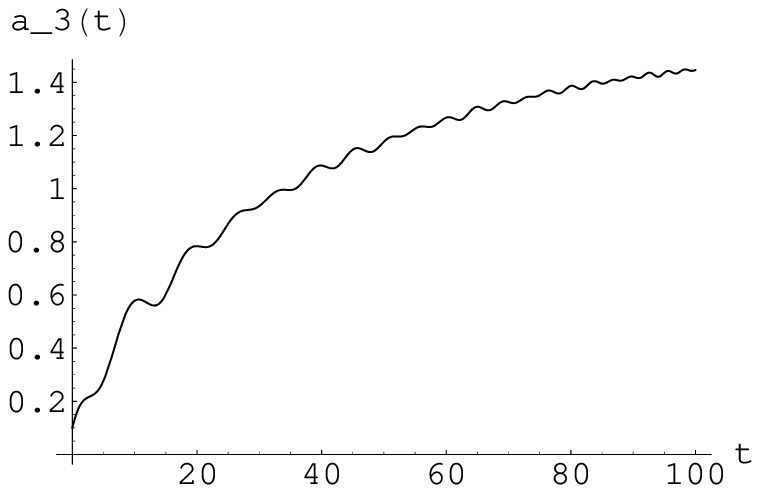}
\end{minipage}
\begin{minipage}{0.33\hsize}
\includegraphics[width=\hsize,keepaspectratio,clip]{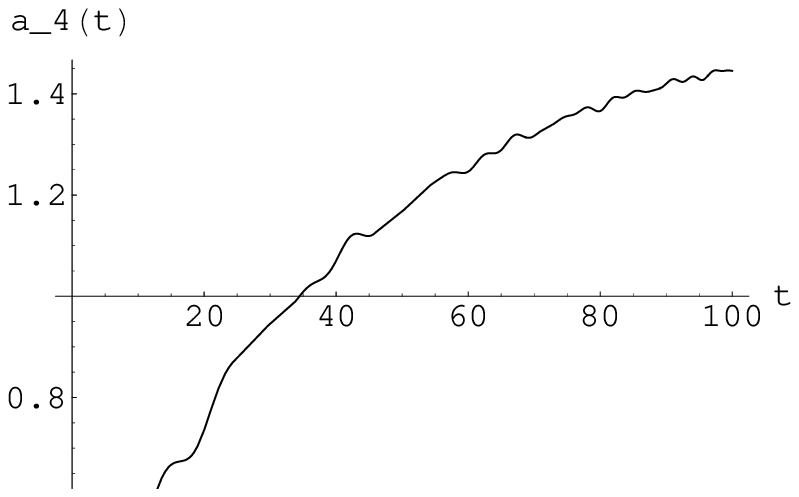}
\end{minipage}
\begin{center}
\caption{A numerical solutions of the scale factors the case of the four-site path graph. }
\end{center}
\end{figure}


\begin{thebibliography}{99}
\bibitem{ref1} T. Hanada, K. Shinoda and K. Shiraishi, 
to appear in Proceedings of the Seventeenth Workshop 
on General Relativity and Gravitation 
(Nagoya University, Nagoya, December 2-6, 2007), arXiv:0801.2641[gr-qc].
\bibitem{ref3.6} N. Kan and K. Shiraishi, 
Class. Quant. Grav. {\bf 20} (2003) 4965 [arXiv:gr-qc/0212113].
\bibitem{MG}   
N.~Arkani-Hamed, H.~Georgi and M.~D.~Schwartz,
Ann. Phys. {\bf 305} (2003) 96 [arXiv:hep-th/0210184];
N.~Arkani-Hamed and M.~D.~Schwartz,
Phys. Rev. {\bf D69} (2004) 104001 [arXiv:hep-th/0302110];
M.~D.~Schwartz,
Phys. Rev. {\bf D68} (2003) 024029 [arXiv:hep-th/0303114];
G.~Cognola, E.~Elizalde, S.~Nojiri, S.~D.~Odintsov and S.~Zerbini,
Mod. Phys. Lett. {\bf A19} (2004) 1435 [arXiv:hep-th/0312269];
S.~Nojiri and  S.~D.~Odintsov,
Phys. Lett. {\bf B590} (2004) 295 [arXiv:hep-th/0403162];
F.~Bauer, T.~Hallgren and G.~Seidl,
Nucl. Phys. {\bf B781} (2007) 32 [arXiv:hep-th/0608176];
G.~Seidl,
e-Print: arXiv:0901.4304 [hep-th].
\bibitem{ref1.5} M. Fierz and W. Pauli, Proc. Roy. Soc. Lond. {\bf A173}
(1939) 211.
\bibitem{ref2} N. Arkani-Hamed, A. G. Cohen and H. Georgi, 
Phys. Rev. Lett. {\bf 86} (2001) 4757 [arXiv:hep-th/0104005]; 
Phys. Lett. {\bf B513} (2001) 232 [arXiv:hep-ph/0105239].
\bibitem{ref3} C. T. Hill, S. Pokorski and J. Wang, 
Phys. Rev. {\bf D64} (2001) 1050050 [arXiv:hep-th/0104035].
\bibitem{ref3.5} S. Hamamoto, 
Prog. Theor. Phys. {\bf 97} (1997) 327 [arXiv:hep-th/9611141].
\bibitem{ref4} N. Kan and K. Shiraishi, 
J. Math. Phys. {\bf 46} (2005) 112301 [arXiv:hep-th/0409268].
\bibitem{ref5} S. G. Nibbelink, M. Peloso and M. Sexton, 
Eur. Phys. J. {\bf C51} (2007) 741 [arXiv:hep-th/0610169].
\bibitem{ref6} S. G. Nibbelink and M. Peloso, 
Class. Quant. Grav. {\bf 22} (2005) 1313 [arXiv:hep-th/0411184].
\bibitem{ref7} C. Bizdadea et al., 
JHEP {\bf 02} (2005) 016; C. Bizdadea et al., Eur. Phys. J. {\bf C48}
(2006) 265.
\end{thebibliography}
\end{document}